\documentclass[12pt]{article}
\usepackage{graphicx,epsfig}
\def\lsim{\mathrel{\raise.3ex\hbox{$<$\kern-.75em\lower1ex\hbox{$\sim$}}}}
\def\gsim{\mathrel{\raise.3ex\hbox{$>$\kern-.75em\lower1ex\hbox{$\sim$}}}}
\newcommand{ \slashchar }[1]{\setbox0=\hbox{$#1$}   
   \dimen0=\wd0                                     
   \setbox1=\hbox{/} \dimen1=\wd1                   
   \ifdim\dimen0>\dimen1                            
      \rlap{\hbox to \dimen0{\hfil/\hfil}}          
      #1                                            
   \else                                            
      \rlap{\hbox to \dimen1{\hfil$#1$\hfil}}       
      /                                             
   \fi}                                             %

\def\to{\rightarrow}

\def\be{\begin{equation}}
\def\ee{\end{equation}}
\def\bea{\begin{eqnarray}}
\def\eea{\end{eqnarray}}
\def\bec{\begin{center}}
\def\eec{\end{center}}
\def\atversim#1#2{\lower0.7ex\vbox{\baselineskip\zatskip\lineskip\zatskip
  \lineskiplimit 0pt\ialign{$\matth#1\hfil##\hfil$\crcr#2\crcr\sim\crcr}}}

\renewcommand{\thefootnote}{\fnsymbol{footnote}}

\newcounter{appendixc}
\newcounter{subappendixc}[appendixc]
\newcounter{subsubappendixc}[subappendixc]

\renewcommand{\appendix}[1] {\vspace*{0.6cm}
        \refstepcounter{appendixc}
        \setcounter{figure}{0}
        \setcounter{table}{0}
        \setcounter{equation}{0}
        \renewcommand{\thefigure}{\Alph{appendixc}.\arabic{figure}}
        \renewcommand{\thetable}{\Alph{appendixc}.\arabic{table}}
        \renewcommand{\theappendixc}{\Alph{appendixc}}
        \renewcommand{\theequation}{\Alph{appendixc}.\arabic{equation}}
        \noindent{\bf Appendix \theappendixc #1}\par\vspace*{0.4cm}}

\begin{document}
\begin{titlepage}
\rightline{\vbox{\halign{&#\hfil\cr &&GSCAS-SPS-05-04 \cr
&&hep-ph/0504087\cr }}} \vskip .5in
\begin{center}

{\Large\bf $D_s^+ - D_s^-$ Asymmetry in Photoproduction}

\vskip .5in \normalsize {\bf  Gang
Hao}\footnote{Email:hao$_-$gang@mails.gscas.ac.cn}$^{2}$, {\bf
Lin Li}\footnote{Email:lilin@gscas.ac.cn}$^{2}$,
{\bf Cong-Feng Qiao}\footnote{Email:cfqiao@gscas.ac.cn}$^{1,2}$\\
\vskip .5cm

$^1$ CCAST(World Lab.), P.O. Box 8730, Beijing 100080, China\\
\vskip .3cm

$^2$ Dept. of Physics, Graduate School of the Chinese
Academy of Sciences,\\
YuQuan Road 19A, Beijing 100049, China \vskip .3cm

\vskip 2.3cm

\end{center}

\begin{abstract}
\normalsize Considering of the possible difference in strange and
antistrange quark distributions inside nucleon, we investigate the
$D_s^+ - D_s^-$ asymmetry in photoproduction in the framework of
heavy-quark recombination mechanism. We adopt two distribution
models of strange sea, those are the light-cone meson-baryon
fluctuation model and the effective chiral quark model. Our
results show that the asymmetry induced by the strange quark
distributions is distinct, which is measurable in experiments.
And, there are evident differences between the predictions of our
calculation and previous estimation. Therefore, the experimental
measurements on the $D_s^+ - D_s^-$ asymmetry may impose a unique
restriction on the strange-antistrange distribution asymmetry
models.
\end{abstract}
\vspace{1cm} PACS number(s):  12.38.Lg, 12.39.Hg, 13.60.Le

\renewcommand{\thefootnote}{\arabic{footnote}}
\end{titlepage}


High energy charmed hadron production plays an important role in
studying strong interactions. Due to the large charm quark mass,
the charm quark involving processes can often be factorized, i.e.,
into a hard process, which describes the interaction scale much
larger than $\Lambda_{QCD}$, the typical QCD renormalization
scale; and a soft part, which demonstrates the hadronization
effects. In recent years, a large production asymmetries between
the charmed and anti-charmed mesons have been measured in
fixed-target hadro- and photo-production experiments
\cite{ex1,ex2,ex3,ex4,cha1,cha2,cha3}. Among these,
photoproduction is thought to give more clean signature than the
hadroproduction ones, because there is only one hadron in the
initial state. However, it is intriguing to notice that the
experimental observation on the asymmetries of the charmed hadron
production are greatly in excess of the predictions of
perturbative QCD(pQCD) \cite{nlo1,nlo2,nlo3,nlo4}.

For large transverse momentum processes, the QCD factorization
theorem \cite{fac} enables us to calculate the cross sections of
heavy hadron production by the pQCD. To be specific, for the $D_s$
photoproduction, the differential cross section may be expressed
as the convolution of the parton distribution function, the
partonic cross section, and the fragmentation function, like

\begin{equation}
d \sigma[\gamma+N \rightarrow D_s+X]=\sum f_{i/N} \bigotimes d
\hat{\sigma}[ \gamma+i \rightarrow c+ \bar{c}+X] \bigotimes D_{c
\rightarrow D_s}, \label{eq:1}
\end{equation}
where $f_{i/N}$is the distribution function of the parton $i$ in a
nucleon $N$; $d \hat{\sigma}( \gamma+i \rightarrow c+ \bar{c}+X)$
is the pQCD calculable cross section of partonic subprocess; and
$D_{c \rightarrow D_s}$ is the nonperturbative fragmentation
function. In this picture, $c$ and $\bar{c}$ are produced
symmetrically at leading order in $\alpha_s$. The asymmetry
appears only in the next-to-leading order (NLO), or higher,
corrections. However, the charm-anticharm asymmetries predicted by
NLO correction are an order of magnitude smaller than the
asymmetries observed in photoproduction experiments\cite{Cuautle}.
That is, the charmed and anticharmed photoproduction asymmetry may
not be fully explained by the charm-anticharm quark production
asymmetry. Moreover, the theoretical predictions through above
mechanism (\ref{eq:1}) cannot account for the differences among
the asymmetries of $D$ mesons with different light quark flavors.

There have already been some early attempts to explain the
observed asymmetries \cite{mod1,mod2}. In these approaches, the
asymmetry is supposed to appear due to the nonperturbative
hadronization effects. Hence, they are all sensitive to unknown
distribution functions of partons in the remnant of the target
nucleon or photon after the hard scattering. In comparison, the
heavy-quark recombination mechanism proposed by Braaten {\it et
al}.\cite{Braaten} can give a more reasonable explanation to the
$D$ meson and $\Lambda_c$ production asymmetries. In the
heavy-quark recombination mechanism, a light parton($q$) that
participates in the hard-scattering process recombines with a
heavy quark($c$) or an antiquark($\bar{c}$) and subsequently
hadronize into the final-state heavy-light meson. The
recombination happens only when the light-quark in the final state
has momentum of $O(\Lambda_{QCD})$ in the heavy-quark, or
antiquark, rest frame. Namely, the light-quark and the heavy-quark
recombine in a small phase space. By virtue of the heavy quark
symmetry, $SU(3)$ flavor symmetry and the large $N_c$ limit of
QCD, the heavy-quark recombination mechanism gives a simple and
predicative explanation for the asymmtries with two
nonperturbative parameters. Since the light quark $q$ in the
recombination model can be either $u$, $d$ or $s$ quark, it can
account for the difference of asymmetries among different light
flavors.

In Ref. \cite{Braatenc}, Braaten {\it et al}. calculated the
charm/anticharm production ratio and asymmetry of $D$ mesons using
the heavy-quark recombination mechanism and confronted their
results to the experimental data. In their consideration, the
asymmetry of $D_s$ meson comes from the process in which the
$\bar{c}(c)$ and light valence-quark of nucleon recombine into a
$D^-(D^+)$ meson, while the recoiling $c(\bar{c})$ quark fragments
to $D_s^+(D_s^-)$ meson. That is, the $D_s^+ - D_s^-$ asymmetry
stems from the excess of $u$ and $d$ over $\bar{u}$ and $\bar{d}$
in the nucleon. Because the $s$ and $\bar{s}$ content of the
nucleon are identical in their assumption, the asymmetry of $D_s$
meson has the opposite sign as that of $D^+$ and $D^-$ meson and
is relatively small. The experimental data on $D_s$ exist
\cite{cha1,cha3}, but with very large errors, and hence do not
tell whether there is an asymmetry or not. Different from their
consideration, recent years, many studies show that there is
striking strange/antistrange sea asymmetry in the momentum
distribution inside the nucleon\cite{sea1,sea2}. Stimulated by
this idea, in this work, we calculate the $D_s$ meson production
asymmetry induced by the strange-antistrange quark distribution
asymmetry within nucleon by employing the heavy-quark
recombination mechanism.

According to the heavy-quark recombination mechanism,  the cross
section of $D_s$ meson photoproduction may schematically
formulated as:
\begin{equation}
d \sigma[\gamma+N \rightarrow D_s +X]=f_{q/N} \otimes \sum d
\hat{\sigma}[\gamma+\bar{s} \rightarrow (c \bar{s} )^n
+\bar{c}]\rho [(c \bar{s} )^n \rightarrow D_s],
\end{equation}
where $(\bar{c} s)^n$ represents that the $s$ in the final state
has small relative momentum in the $\bar{c}$ rest frame, and $n$
is the color and angular momentum quantum numbers of $(\bar{c} s)$
intermediate state. $d \hat{\sigma}[\gamma+ \bar{s} \rightarrow (c
\bar{s})^n +\bar{c}]$ is the perturbative QCD calculable partonic
subprocess. The factor $\rho[(c \bar{s})^n \rightarrow D_s]$ is
the probability of the $(\bar{c} s)^n$ state to evolve into a
final state, here, the $D_s$.

In our consideration the $D_s$ meson may be produced via two
different schemes, i.e.
\begin{equation}
(a)~~ d \hat{\sigma}[\gamma+ \bar{s} \rightarrow (c \bar{s})^n
+\bar{c}]\rho [(c \bar{s})^n \rightarrow D_s]\;, \label{recom}
\end{equation}
\begin{equation}
(b)~~ d \hat{\sigma}[\gamma+ q \rightarrow (\bar{c} q)^n
+c]\sum_{\bar D} \rho [(\bar c q)^n \to \bar{D}] \otimes D_{c
\rightarrow D_s}\; , \label{recom-frag}
\end{equation}
while in Ref.\cite{Braatenc} only the second one, the
recombination-fragmentation mechanism, was taken into account. In
process (a), the $(c \bar{s})^n$ recombines into the $D_s$ meson
directly; in process (b), $(\bar{c} q)^n$ recombines into the
$D_s^-$, $D^-$ or $\bar{D}^0$ meson, and the recoiling $c$ quark
fragments to the $D_s^+$ meson.

The calculation of the partonic cross section $d \hat{\sigma}[\gamma+q
\rightarrow (\bar{c} q)^n +c]$ in pQCD is straightforward, and our calculation
confirms the Ref.\cite{Braatenc} results. That is,
\begin{eqnarray}
\frac{d \hat{\sigma}}{dt}[\bar{c}q(^1 S^{(1)}_0)]&=& \frac{256
\pi^2 e^2_c \alpha \alpha^2_s}{81} \frac{m^2_c}{S^3}[-
\frac{S}{U}(1+ \frac{\kappa T}{S})^2
 \nonumber \\
&+& \frac{m^2_cS}{U^2}(- \frac{S^3}{T^3}+ \frac{2(1+ \kappa)S}{T}
+4 \kappa + \frac{\kappa^2 T}{S})+\frac{2m^4_c S^3}{T^3 U^2}(1+
\frac{\kappa T}{S})]\; ,
\end{eqnarray}

\begin{eqnarray}
\frac{d \hat{\sigma}}{dt}[\bar{c}q(^3 S^{(1)}_1)]&=& \frac{256
\pi^2 e^2_c \alpha \alpha^2_s}{81} \frac{m^2_c}{S^3}[-
\frac{S}{U}(1+ \frac{2 U^2}{T^2})(1+ \frac{\kappa T}{S})^2
 \nonumber \\
&+& \frac{m^2_cS}{U^2}(\frac{S^3}{T^3}+ \frac{4(2+
\kappa)S^2}{T^2} +\frac{2(3+ 7\kappa)S}{T}+4 \kappa (3+\kappa) +
\frac{ 3 \kappa^2 T}{S})  \nonumber \\
&+&\frac{6 m^4_c S^3}{T^3 U^2}(1+ \frac{\kappa T}{S})]\; .
\end{eqnarray}
Here, $\kappa=e_q/e_c$ is the ratio of the electric charge fractions
of light and charm quarks. The Lorentz invariants are defined as
$S=(p_q + p_\gamma)^2$, $T=(p_\gamma -p_c)^2-m_c^2$ and $U=(p_\gamma
-p_{\bar{c}})^2-m_c^2$. $p_q$, $p_\gamma$ and $p_c$ are the
momenta of the light quark, photon, and $c$ quark, respectively.
It is noted that because the relatively small momentum of the light
quark in $(c \bar{s})^n$ system, the higher angular momentum excited
states are suppressed by power of $\Lambda_{QCD}/p_T$ or $\Lambda_{QCD}/m_c$.
For a leading order estimation, we take only the contributions from
$^1 S_0$ and $^3S_1$ states.

To calculate the total cross section ($d \sigma[\gamma+N
\rightarrow D_s +X]$) of the production of $D_s$ meson, we need to
know the distribution of strange sea quark. According to the
light-cone meson-baryon fluctuation model\cite{sea1}, the strange
quark-antiquark asymmetry of the nucleon sea is generated by the
intrinsic strangeness fluctuations in the proton wavefunction. In
this model, the asymmetry stems from the intermediate $K^+\; \Lambda$
configuration of the incident nucleon, which has the lowest
off-shell light-cone energy and invariant mass \cite{klambda}. The
momentum distributions of the intrinsic strange and antistrange
quarks in the $K^+\; \Lambda$ system can be parameterized as
\begin{eqnarray}
s(x)=\int^{1}_{x}~\frac{dy}{y}f_{\Lambda/K^+ \Lambda}(y)
q_{s/\Lambda}(\frac{x}{y})\; ,
\end{eqnarray}
\begin{eqnarray}
\bar{s}(x)=\int^{1}_{x}~\frac{dy}{y}f_{K^+/K^+ \Lambda}(y)
q_{\bar{s}/K^+}(\frac{x}{y})\; ,
\end{eqnarray}
where $f_{\Lambda/K^+ \Lambda}(y)$, $f_{K^+/K^+ \Lambda}(y)$ are
the probabilities of finding $\Lambda$, $K^+$ in the ($K^+ \Lambda$)
state; $q_{s/\Lambda}(\frac{x}{y})$,
$q_{\bar{s}/K^+}(\frac{x}{y})$ are probabilities of finding
strange and antistrange quark in $\Lambda$ or $K^+$ states. To
estimate these quantities, two simple functions of the invariant
mass $M^2=\Sigma^2_{i=1}~\frac{K^2_{\bot i} +m^2_i}{x_i}$ for the
two-body wavefunction are given \cite{wave},
\begin{eqnarray}
\psi_{Gaussian}(M^2)=A_{Gaussian}exp(-M^2/2 \alpha^2)\; ,
\end{eqnarray}
\begin{eqnarray}
\psi_{Power}(M^2)=A_{power}(1+M^2/\alpha^2)^{-p}\; ,
\end{eqnarray}
where the $\alpha$ represents the typical internal momentum scale.
In our analysis in this work, we simply adopt the Gaussian type
wavefunction for use.

In recently, another model, which based on the effective chiral quark
theory, for the distribution of strange sea quark
is proposed by Ma {\it et al}. \cite{sea2}. In this model, the
strange quark distribution is determined by both the constituent quark
distribution and the quark splitting function. For instance,
\begin{eqnarray}
s(x)=P_{sK^+ /u} \bigotimes u_{0}+ P_{sK^0 /d} \bigotimes d_{0}\; ,
\end{eqnarray}
and
\begin{eqnarray}
\bar{s}(x)=V_{\bar{s} /K^+}  \bigotimes P_{K^+ s/u} \bigotimes
u_{0}+ V_{\bar{s} /K^0} \bigotimes P_{K^0 s /d} \bigotimes d_{0}\; .
\end{eqnarray}
Here, $P_{j \alpha s/i}$ is the splitting function representing the
probability of finding a constituent quark $j$ together with a
spectator Goldstone(GS) boson ($\alpha= \pi, K, \eta$) in a parent
constituent quark. $u_{0}$ and $d_{0}$ denote the constituent quark
distributions; $V_{i/\alpha}$ is the quark $i$ distribution within
the GS boson $\alpha$.

Using the resultant production cross section, it is
straightforward to calculate the asymmetry of the $D_s^+ - D_s^-$
photoproduction, which are defined as
\begin{eqnarray}
\alpha[D_s]&=& \frac{\sigma_{D_s^+}-
\sigma_{D_s^-}}{\sigma_{D_s^+}+ \sigma_{D_s^-}}\; .
\end{eqnarray}
In our calculation, the two nonperturbative input parameters,
$\rho_{sm}$ and $\rho_{sf}$, are extracted from experiments by
fiting to the E687 and E691 data \cite{cha1,cha3}.
\begin{eqnarray}
\rho_{sm}&=&\rho_{eff}[c \bar{u}(^1 S_0) \rightarrow
D^0]=\rho_{eff}[c \bar{d}(^1 S_0) \rightarrow D^+]=\rho_{eff}[c
\bar{s}(^1 S_0) \rightarrow D^+_s] \nonumber \\
&=&\rho_{eff}[c \bar{u}(^3 S_1) \rightarrow D^{*0}]=\rho_{eff}[c
\bar{d}(^3 S_1) \rightarrow D^{*+}]=\rho_{eff}[c \bar{s}(^3 S_1)
\rightarrow D^{*+}_s]=0.15\; , \nonumber \\
\end{eqnarray}
and
\begin{eqnarray}
\rho_{sf}&=&\rho_{eff}[c \bar{u}(^3 S_1) \rightarrow
D^0]=\rho_{eff}[c \bar{d}(^3 S_1) \rightarrow D^+]=\rho_{eff}[c
\bar{s}(^3 S_1) \rightarrow D^+_s] \nonumber \\
&=&\rho_{eff}[c \bar{u}(^1 S_0) \rightarrow D^{*0}]=\rho_{eff}[c
\bar{d}(^1 S_0) \rightarrow D^{*+}]=\rho_{eff}[c \bar{s}(^1 S_0)
\rightarrow D^{*+}_s]=0\; .\nonumber \\
\end{eqnarray}
Here, the subscripts $sm$ and $sf$ stand for spin-matched and
spin-flipped situations, respectively. Considering the possible
uncertainties, the above values are in agreement with those
extracted from the hadroproduction \cite{hadron}.

As for the fragmentation function of charm quark to the $D_s$ meson,
we exploit the well-known Peterson fragmentation function,
\begin{eqnarray}
D_{c \rightarrow D_s} (z) = P(z;\epsilon)\; f_{c\rightarrow D_s}\; .
\end{eqnarray}
Here, $D(z;\epsilon)$ \cite{peter} is the Peterson function and
$f_{c \rightarrow D_s}$\cite{frag} is the fragmentation
probability. In practice, the parameter $\epsilon$ is set to be
$0.06$, as in Ref. \cite{Braatenc}, which falls in the region of
experimental fitting allowance. All through this paper, the charm
quark mass are set to be $1.5$ GeV, and the factorization and
renormalization scales are set to be at the meson transverse mass
$m_T$, the $\sqrt{p^2_{\bot} + m^2_{D_s}}$. The parton
distribution distribution function(PDF) of CTEQ6L \cite{cte6} is
used. For simplicity, we use the average photon beam energy of the
E687 experiment, $\langle E_\gamma \rangle =200GeV$, in the
calculation. To be more accurate, the $D_s^*$ feed down effect is
included in our analysis, but those from other higher excited
states are thought to be small and simply neglected.

In figures 1 and 2, based on the light-cone meson-baryon fluctuation
model, the dependence of the inclusive cross sections of $D^+ _s$ and
$D_s ^-$ production on the Feynman variable $x_F$ and the rapidity $y$
respectively are shown. The corresponding effective chiral quark model
results are presented in Figures 3 and 4.
\begin{figure} \vspace{0.0cm}
\begin{center}
\includegraphics[scale=1.]{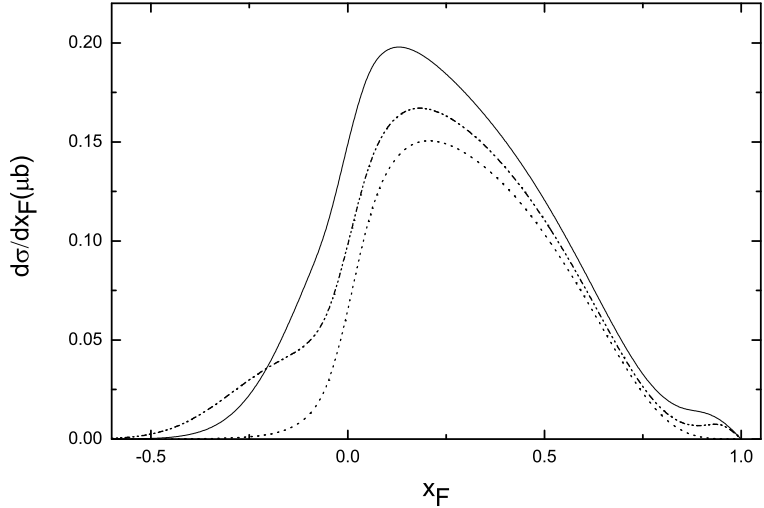} \vspace{-1.cm}
\caption{The inclusive cross sections $d \sigma/dx_F$, calculated by
using the light-cone meson-baryon fluctuation model, for $D_s$ production
with $\rho_{sm} = 0.15$ and $\rho_{sf} = 0$. The solid, dash-dotted, and
dotted lines correspond to the production of $D_s^+$, $D_s^-$ in
recombination mechanism, and $D_s^ \pm$ at the leading order photon-gluon
interaction with the fragmentation mechanism,
respectively.}
\end{center} \label{block1}
\end{figure}
\begin{figure} \vspace{0.0cm}
\begin{center}
\includegraphics[scale=1.]{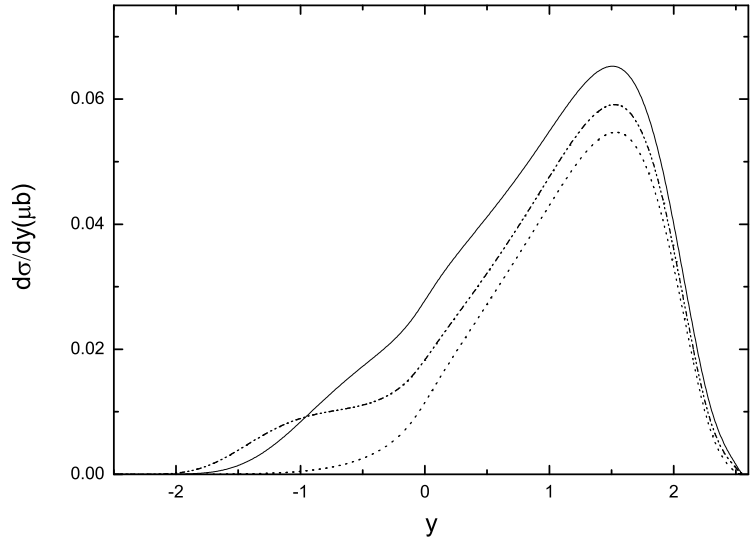} \vspace{-1.cm}
\caption{Inclusive cross sections $d \sigma/dy$ calculated with
the light-cone meson-baryon fluctuation model. The curves are described
as in Figure 1. }
\end{center} \label{block1}
\end{figure}
\begin{figure} \vspace{0.0cm}
\begin{center}
\includegraphics[scale=1.]{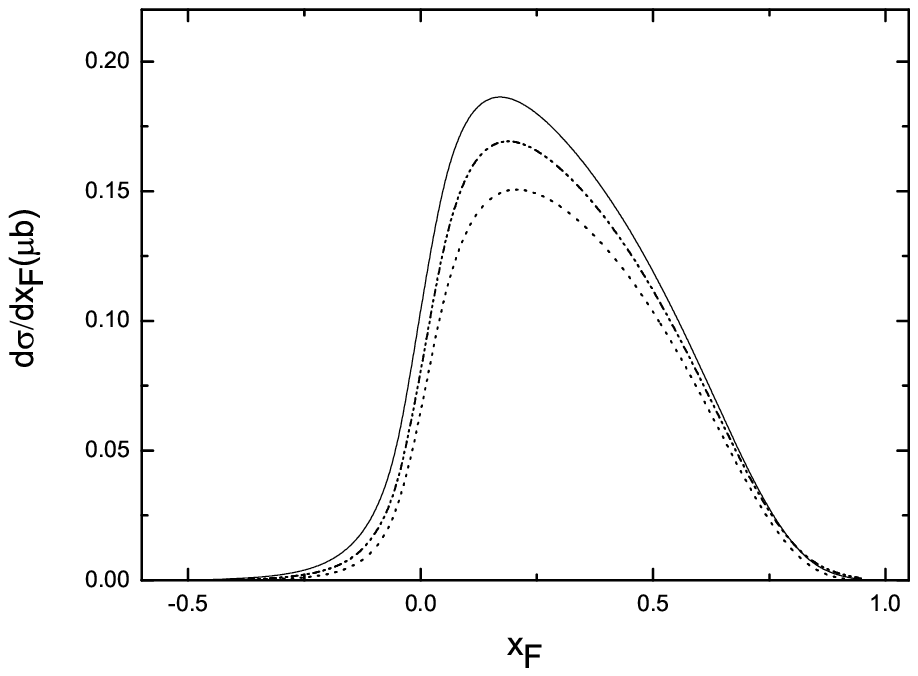} \vspace{-1.cm}
\caption{Inclusive cross sections $d \sigma/dx_F$ calculated with
the effective chiral quark model. The curves are described
as in Figure 1.}
\end{center} \label{block1}
\end{figure}
\begin{figure} \vspace{0.0cm}
\begin{center}
\includegraphics[scale=1.]{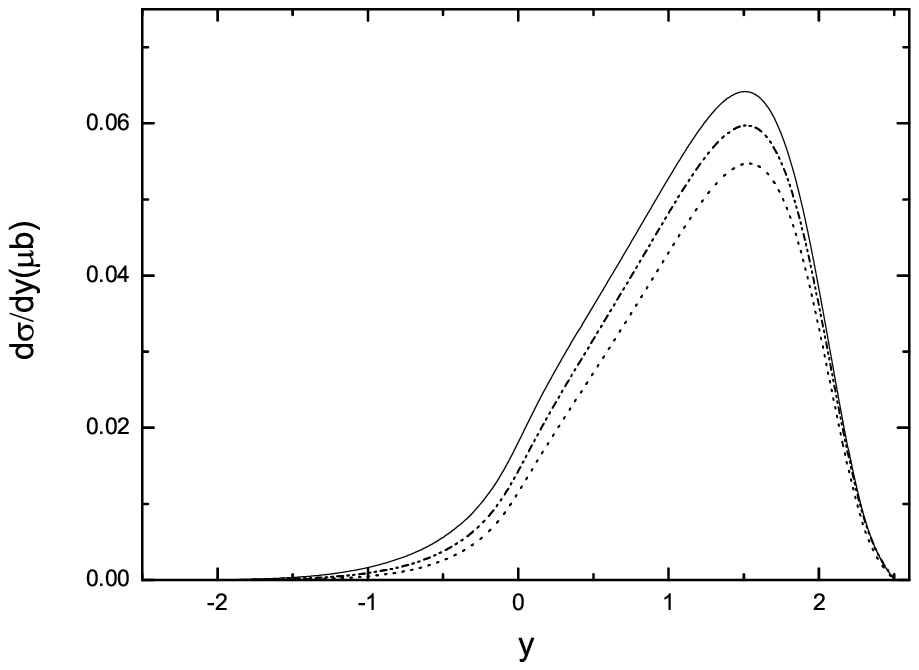} \vspace{-1.cm}
\caption{Inclusive cross sections $d \sigma/dx_y$ calculated with
the effective chiral quark model. The curves are described
as in Figure 1.}
\end{center} \label{block1}
\end{figure}
For comparison, the cross section of the leading order
photon-gluon fusion process, which will give a symmetric
$D_s^{\pm}$ production through fragmentation, is also presented.
We find that the results appear to be insensitive to the
variations of the factorization scale and the Peterson parameter
$\epsilon$. From Figs. 1 - 4, we see that cross sections peak in
the region of small $x_F$ and large rapidity $y$. Although the
contributions from the recombination mechanism are not the
dominant ones, they may give out the asymmetry. Both for the
light-cone meson-baryon fluctuation model and the effective chiral
quark model, in the region of the experimental cut $x_F > 0$, the
corresponding cross sections for $D_s^+$ are bigger than those for
the $D_s^-$. Thus, the $D_s$ photoproduction asymmetry appears with
the effect of strange-antistrange quark asymmetric distributions.
\begin{figure} \vspace{0.0cm}
\begin{center}
\includegraphics[scale=1.]{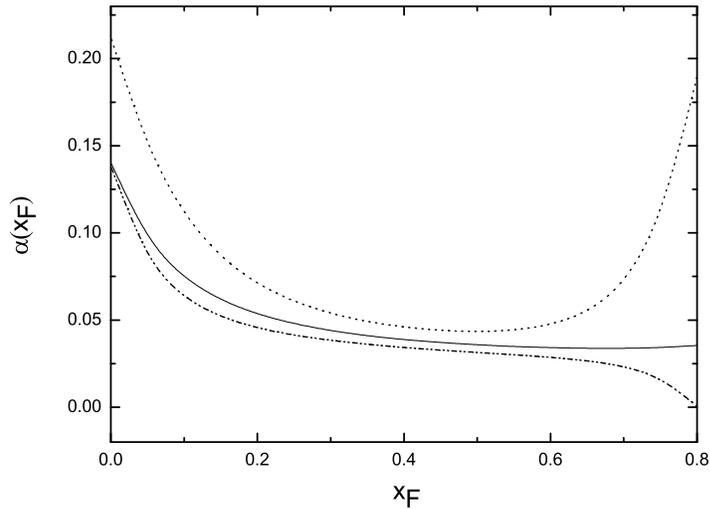} \vspace{-1.cm}
\caption{The asymmetry $\alpha[D_s]$ versus $x_F$. The dotted and
dash-dotted lines correspond to the results from the light-cone
meson-baryon fluctuation model and the effective chiral
quark model, respectively. The solid line from the
Ref.\cite{Braatenc} result. }
\end{center} \label{block1}
\end{figure}

Our predicted $D_s$ production asymmetries, in comparison with
Ref.\cite{Braatenc}, are shown in in Figure 5. We find that, by
adopting the light-cone meson-baryon fluctuation model, the
production asymmetry of $D_s$ meson is about 1.2 times larger than
that in Ref.\cite{Braatenc}; and by using the effective chiral
quark model, it is about $80 \%$ bigger only. If we extend to the
$x_F$ negative region, the asymmetries coming from the intrinsic
strange sea momentum distribution are very big and diverge very
much from the prediction of Ref.\cite{Braatenc}. In this region,
the asymmetry from the light-cone meson-baryon fluctuation model
is even flipped relative to the positive $x_F$ region, while the
effective chiral quark model and the fragmentation-recombination
scheme give the results with the same sign as in the $x_F > 0$
region. Since the obtained data are in the
$x_F > 0$ region, here we will not show the results in the $x_F <
0 $ region. From the figure 5, it is obvious that the three
different asymmetry producing schemes(models) give results
diverging with each other with the $x_F$ increase. This scaling
difference leaves the experiments with an opportunity to decide
the physical reality.

In summary, in this paper, the heavy-quark
recombination mechanism was employed, which gets a first success in
explaining the $D$ meson and $\Lambda_c$ asymmetry production.
We have studied the $D_s^+ - D_s^-$
asymmetry in the photoproduction. Our point is that this
asymmetry can be induced not only by the recombination-fragmentation
mechanism, the Eq.\ref{recom-frag}, but also by the strange and
antistrange distribution asymmetry inside the nucleon, the Eq.\ref{recom}.
And, we find the latter effect is even bigger then the former one
depending on the employed model and phase space region.
Two QCD relevant strange quark distribution models were used, that is
the light-cone meson-baryon fluctuation model and the effective
chiral quark model. After including the process (\ref{recom}),
the predicted asymmetry increases anyhow. However,
what we discussed should be the dominant contributions to
the $D_s$ photoproduction asymmetry.
Although there remains some uncertainties, such as from the
breakdown of $SU(3)$ symmetry, the use of large $N_c$ limit and the
NLO corrections to the leading order photon-gluon fusion, etc.,
our results are adequate to make qualitative predictions, and more
detailed discussion on the relevant uncertainties can be found
in Ref.\cite{Braaten,Braatenc}. It is noticed that
the nowadays $D_s$ asymmetry experimental data are very limited and
preliminary. With enough data collection in the future, it is expected
that the experiment can inversely impose a strong restriction on
the strange and antistrange quark distributions by measuring
the $D_s$ production asymmetry.

\vspace{20mm} \noindent {\Large\bf Acknowledgements:} We thank
Y.Jia and B.Q. Ma for helpful discussions on this topic. This work
was supported in part by the Natural Science Foundation of China (NSFC).\\

\end{document}